\documentclass[a4paper]{jpconf}
\usepackage{graphicx}
\usepackage{amsmath,amsfonts,amssymb} 
\usepackage[english]{babel} 
\usepackage{hyperref} 
\usepackage{mathtools} 
\usepackage{microtype} 
\usepackage[capitalise]{cleveref} 
\usepackage{listings}
\usepackage{color}
\usepackage{xspace}

\newcommand{\feyncalc}{\textsc{FeynCalc}\xspace}
\newcommand{\feynarts}{\textsc{FeynArts}\xspace}
\newcommand{\mathematica}{\textsc{Mathematica}\xspace}

\begin{document}
\title{FeynCalc 9}

\author{Vladyslav Shtabovenko}

\address{Technische Universit\"at M\"unchen, Physik-Department T30f, James-Franck-Str. 1, 85747 Garching, Germany}

\ead{v.shtabovenko@tum.de}

\begin{abstract}
We report on a new version of \feyncalc, a well-known \mathematica package for symbolic computations in quantum field theory and provide some explicit examples for using the software in different types of calculations.
\end{abstract}

\section{Introduction}

Modern perturbative calculations in quantum field theory (QFT) often heavily rely on different software tools that are used to automatize both the algebraic and the numerical stages of the evaluation.

In the last decades, much effort was invested to develop packages (e.g. \textsc{FormCalc} \cite{Hahn1999}, \textsc{GoSam} \cite{Cullen2014}, \textsc{GRACE} \cite{Belanger2006}) that are  able to perform all the steps needed to arrive to physical observables (e.g. decay rates or cross-sections) in a highly automatized way.
The level of the automation in such packages is so advanced, that in many cases the
user just has to specify the incoming and outgoing particles, while the underlying
code will take care of everything else and produce final results.

A complementary approach is represented by tools (e.g. \textsc{HEPMath} \cite{Wiebusch2014}, \textsc{Package-X} \cite{Patel2015}) that do not attempt to automatize everything bur rather provide a collection of easy to use instruments for accomplishing most common tasks in QFT calculations, like contractions of Lorentz indices, calculation of Dirac traces or tensor decomposition of loop integrals. Such semi-automatic packages give the user much flexibility but also require him or her to carefully handle every step of the calculation. This bears a strong resemblance to calculations done by pen and paper, with the difference that one can proceed in a much faster way .

While both philosophies have their supporters in the high energy physics community, one should of course understand that they are not competing with each other but rather cover different use cases. On the one hand, many tree-level and 1-loop calculations in Standard Model and its popular (e.g. supersymmetric) extensions can be nowadays carried out in a highly automatic fashion, since the essential steps of these calculations are sufficiently well understood. It is therefore natural to use suitable packages like
\textsc{FormCalc} or \textsc{GoSam} to study such processes.

On the other hand, multi-loop calculations, determination of matching coefficients in effective field theories or evaluation of Feynman diagrams in non-relativistic QFTs are good examples for tasks that are too complex or non-standard to be automatized in full generality.
Although for such projects people often prefer to write their own code from scratch (e.g. implemented in \textsc{FORM} \cite{Vermaseren2007} or \mathematica), it can also be useful to use different instruments provided by publicly available semi-automatic packages instead of or in addition to the private codes.

\feyncalc belongs to the oldest semi-automatic packages that are still actively developed and used in research. The first
version of the program was published almost 25 years ago \cite{Mertig1991} and followed  the idea to provide a flexible and easy to use \mathematica package for symbolic calculations and for the evaluation of Feynman diagrams in QFT at tree-level and at 1-loop. In this context,
"symbolic calculations" means that \feyncalc can deal not only with Feynman diagrams but also standalone QFT expressions  like
$\int \frac{d^D q}{(2\pi)^D} \frac{l^\mu l^\nu}{l^2-m^2}$ or $\Tr(\gamma^\mu \gamma^\nu \gamma^\rho \gamma^\sigma)$. The users of \feyncalc
are therefore not forced to follow a particular workflow (e.g. first enter the Lagrangian, then generate Feynman diagrams and finally evaluate them)
but are free to use the software as some sort of "calculator" for QFT expressions. This design philosophy has been preserved over the years and made \feyncalc a valuable tool for many QFT practitioners.

This note is organized in the following way. In Sec. \ref{sec:features} we briefly describe the most interesting new features of \feyncalc 9 and provide some examples for purposes of illustration. Sec. \ref{sec:gluon} shows the calculation of the gluon self-energy at 1-loop in QCD, where the final result will be presented in terms of the Passarino-Veltman scalar functions $A_0$ and $B_0$. In Sec. \ref{sec:summary} we summarize and mention main directions for the future development of \feyncalc.

\section{New features in \feyncalc 9} \label{sec:features}

\feyncalc 9 was released in January 2016 \cite{Shtabovenko2016}. While the development of the program between 2001 and 2014 was mostly limited to providing fixes for the  discovered bugs,
this version also introduces some new features that are mostly related to the evaluation of 1-loop and multi-loop integrals. Version 9 requires at least \mathematica 8 and can be installed by just evaluating

\begin{lstlisting}[language=Mathematica,mathescape]
Import["https://raw.githubusercontent.com/FeynCalc/feyncalc/master/install.m"]
InstallFeynCalc[]
\end{lstlisting}
in a \mathematica notebook. The automatic installer also handles the setup of \feynarts \cite{Hahn2001}, a \mathematica package that can generate Feynman diagrams, which can be then (after some conversion) evaluated with \feyncalc. An example for the conversion from \feynarts to \feyncalc will be described in Sec. \ref{sec:gluon}.

Tensor decomposition of 1-loop integrals using the \texttt{TID} routine has received a lot of improvements in \feyncalc 9 and can in principle
handle 1-loop tensor integrals of arbitrary rank and multiplicity. Provided that there are no vanishing Gram determinants, one can choose (via the option \texttt{UsePaVeBasis})
between the full decomposition into Passarino-Veltman scalar functions

\begin{lstlisting}[language=Mathematica,mathescape]

In[1]:= int = FCI[GAD[$\mu$].(m + GSD[q]).GAD[$\mu$] FAD[{q, m},{q-p}]]

Out[1]:= $\frac{\gamma ^{\mu }.(m+\gamma \cdot q).\gamma ^{\mu }}{\left(q^2-m^2\right).(q-p)^2}$

In[2]:= TID[int, q]//ToPaVe[#, q]&


Out[2]= $\frac{i \pi ^2 (D-2) \text{A}_0\left(m^2\right) \gamma \cdot p}{2 p^2}$-$\frac{i \pi ^2 \text{B}_0\left(p^2,0,m^2\right) \left(D m^2 \gamma \cdot p-2 D m p^2+D p^2 \gamma \cdot p-2 m^2 \gamma \cdot p-2 p^2 \gamma \cdot p\right)}{2 p^2}$
\end{lstlisting}
and the decomposition into Passarino-Veltman coefficient functions

\begin{lstlisting}[language=Mathematica,mathescape]

In[3]:= TID[int, q, UsePaVeBasis -> True, PaVeAutoReduce -> False] // 
				ToPaVe[#, q] &

Out[3]:= $i \pi ^2 \text{B}_0\left(p^2,0,m^2\right) (D m-D \gamma \cdot p+2 \gamma \cdot p)-i \pi ^2 (D-2) \gamma \cdot p \text{B}_1\left(p^2,0,m^2\right)$
\end{lstlisting}
Zero Gram determinants are automatically recognized and force the algorithm to avoid the full decomposition

\begin{lstlisting}[language=Mathematica,mathescape]

In[4]:= FCClearScalarProducts[]
				SPD[p1, p2] = x;
				SPD[p1, p1] = x;
				SPD[p2, p2] = x;
				FCI[FVD[l, $\mu$] FVD[l, $\nu$] FVD[l, $\rho$] FAD[{l, m0}, {l + p1, m1}, {l + p2, m2}]]

Out[4]:= $\frac{l^{\mu } l^{\nu } l^{\rho }}{\left(l^2-\text{m0}^2\right).\left((l+\text{p1})^2-\text{m1}^2\right).\left((l+\text{p2})^2-\text{m2}^2\right)}$

In[5]:= TID[(1/(I Pi^ 2)) %, l]

Out[6]:= $\left(\text{p1}^{\rho } g^{\mu  \nu }+\text{p1}^{\nu } g^{\mu  \rho }+\text{p1}^{\mu } g^{\nu  \rho }\right) \text{C}_{001}\left(x,0,x,\text{m0}^2,\text{m1}^2,\text{m2}^2\right)$
$+\left(\text{p2}^{\rho } g^{\mu  \nu }+\text{p2}^{\nu } g^{\mu  \rho }+\text{p2}^{\mu } g^{\nu  \rho }\right) \text{C}_{002}\left(x,0,x,\text{m0}^2,\text{m1}^2,\text{m2}^2\right)$
$+\text{p1}^{\mu } \text{p1}^{\nu } \text{p1}^{\rho } \text{C}_{111}\left(x,0,x,\text{m0}^2,\text{m1}^2,\text{m2}^2\right)$
$+\left(\text{p1}^{\nu } \text{p1}^{\rho } \text{p2}^{\mu }+\text{p1}^{\mu } \text{p1}^{\rho } \text{p2}^{\nu }+\text{p1}^{\mu } \text{p1}^{\nu } \text{p2}^{\rho }\right) \text{C}_{112}\left(x,0,x,\text{m0}^2,\text{m1}^2,\text{m2}^2\right)$
$+\left(\text{p1}^{\rho } \text{p2}^{\mu } \text{p2}^{\nu }+\text{p1}^{\nu } \text{p2}^{\mu } \text{p2}^{\rho }+\text{p1}^{\mu } \text{p2}^{\nu } \text{p2}^{\rho }\right) \text{C}_{122}\left(x,0,x,\text{m0}^2,\text{m1}^2,\text{m2}^2\right)$
$+\text{p2}^{\mu } \text{p2}^{\nu } \text{p2}^{\rho } \text{C}_{222}\left(x,0,x,\text{m0}^2,\text{m1}^2,\text{m2}^2\right)$
\end{lstlisting}
where we multiplied the result by $\tfrac{1}{i \pi^2}$ to cancel the normalization factor that appears in the conversion of loop integrals to Passarino-Veltman functions.

With the new function \texttt{FCMultiLoop\-TID} it is now also possible to compute tensor decompositions of multi-loop integrals, although in this case a special treatment of integrals with zero Gram determinants is not available yet.

\begin{lstlisting}[language=Mathematica,mathescape,escapechar=?]
In[7]:= FCI[FVD[q1, $\mu$] FVD[q2, $\nu$] FAD[q1, q2, {q1 - p1}, {q2 - p1}, {q1 - q2}]]

Out[7]:= $\frac{\text{q1}^{\mu } \text{q2}^{\nu }}{\text{q1}^2.\text{q2}^2.(\text{q1}-\text{p1})^2.(\text{q2}-\text{p1})^2.(\text{q1}-\text{q2})^2}$

In[8]:= FCMultiLoopTID[%, {q1, q2}]

Out[8]:= $\frac{D \text{p1}^{\mu } \text{p1}^{\nu }-\text{p1}^2 g^{\mu  \nu }}{4 (D-1) \text{q2}^2.\text{q1}^2.(\text{q2}-\text{p1})^2.(\text{q1}-\text{q2})^2.(\text{q1}-\text{p1})^2}-$

$\frac{\text{p1}^2 g^{\mu  \nu }-\text{p1}^{\mu } \text{p1}^{\nu }}{2 (D-1) \text{p1}^2 \text{q2}^2.\text{q1}^2.(\text{q2}-\text{p1})^2.(\text{q1}-\text{p1})^2}+\frac{\text{p1}^2 g^{\mu  \nu }-\text{p1}^{\mu } \text{p1}^{\nu }}{(D-1) \text{p1}^2 \text{q2}^2.\text{q1}^2.(\text{q1}-\text{q2})^2.(\text{q1}-\text{p1})^2}$-$\frac{D \text{p1}^{\mu } \text{p1}^{\nu }-\text{p1}^2 g^{\mu  \nu }}{2 (D-1) \text{p1}^4 \text{q1}^2.(\text{q2}-\text{p1})^2.(\text{q1}-\text{q2})^2}$
\end{lstlisting}

Simplification of scalar loop integrals with linearly dependent propagators is another feature of \feyncalc that was greatly improved in the version 9 by adopting the partial fractioning algorithm of F. Feng \cite{Feng2012a}. The corresponding
function \textsc{ApartFF} is not only faster and more efficient than the old \textsc{ScalarProductCancel} (which is now considered legacy), but is also capable to handle multi-loop integrals
\begin{lstlisting}[language=Mathematica,mathescape,escapechar=!]
In[9]:= FCI[SPD[p, q1] ^2 SPD[p, q2]  FAD[{q1, m}, {q2, m}, q1 - p, q2 - p,  q1 - q2]]
   
Out[9]:= $\frac{(p  \cdot  \text{q2}) (p \cdot \text{q1} )^2}{\left(\text{q1}^2-m^2\right).\left(\text{q2}^2-m^2\right).(\text{q1}-p)^2.(\text{q2}-p)^2.(\text{q1}-\text{q2})^2}$

In[10]:= ApartFF[%,{q1,q2}]

Out[10]:=$\frac{\left(m^2+p^2\right)^3}{8 \left(\text{q1}^2-m^2\right).\left(\text{q2}^2-m^2\right).(\text{q2}-p)^2.(\text{q1}-\text{q2})^2.(\text{q1}-p)^2}-\frac{\left(m^2+p^2\right)^2}{4 \left(\text{q1}^2-m^2\right).\left(\text{q2}^2-m^2\right).(\text{q1}-\text{q2})^2.(\text{q1}-p)^2}$
$+\frac{\left(m^2+p^2\right)^2}{4 \text{q2}^2.\text{q1}^2.\left((\text{q1}-p)^2-m^2\right).(\text{q1}-\text{q2})^2}+\frac{\left(m^2+p^2\right) \left(p^2-p\cdot \text{q1}\right)}{4 \text{q2}^2.\text{q1}^2.(\text{q1}-\text{q2})^2.\left((\text{q2}-p)^2-m^2\right)}$
$-\frac{\left(m^2+p^2\right) (p\cdot \text{q1})}{4 \left(\text{q1}^2-m^2\right).\left(\text{q2}^2-m^2\right).(\text{q2}-p)^2.(\text{q1}-\text{q2})^2}-\frac{p\cdot \text{q1}}{4 \left(\text{q2}^2-m^2\right).(\text{q1}-\text{q2})^2.(\text{q1}-p)^2}-\frac{m^2+p\cdot \text{q1}+p^2}{4 \left(\text{q1}^2-m^2\right).(\text{q2}-p)^2.(\text{q1}-\text{q2})^2}$
$+\frac{m^2+2 (p\cdot \text{q1})+p^2}{8 \left(\text{q1}^2-m^2\right).\left(\text{q2}^2-m^2\right).(\text{q2}-\text{q1})^2}$

\end{lstlisting}
A more detailed description of these and other new and improved functions can be found in \cite{Shtabovenko2016}. 

\section{Gluon self-enery at 1-loop in QCD} \label{sec:gluon}
To make some connection to more realistic examples, let us look at the 1-loop gluon self-energy in QCD. While this is, of course, a standard textbook calculation that can be done by pen and paper with comparably little effort, it shows how \feyncalc can be used to evaluate Feynman diagrams generated by \feynarts.

The first step is always to load \feyncalc with the global option \texttt{\$LoadFeynArts} set to \texttt{True}, which ensures that the patched copy of \feynarts, that usually resides inside the \feyncalc installation, will we loaded in a proper way. It is also convenient to set the  option \texttt{\$FAVerbose} to \texttt{0}, which prevents \feynarts from displaying too much extra information.

\begin{lstlisting}[language=Mathematica]
In[1]:= $LoadFeynArts = True;
<< FeynCalc`
$FAVerbose = 0;
\end{lstlisting}

Then we can use \feynarts' functions \texttt{CreateToplogies} and \texttt{InsertFields} \footnote{The syntax for these and other \feynarts functions can be found in the "FeynArts User's Guide", available from \url{www.feynarts.de} } to generate the diagrams that we want to evaluate.

\begin{figure}[h!]
\centering
\includegraphics[width=0.7\textwidth]{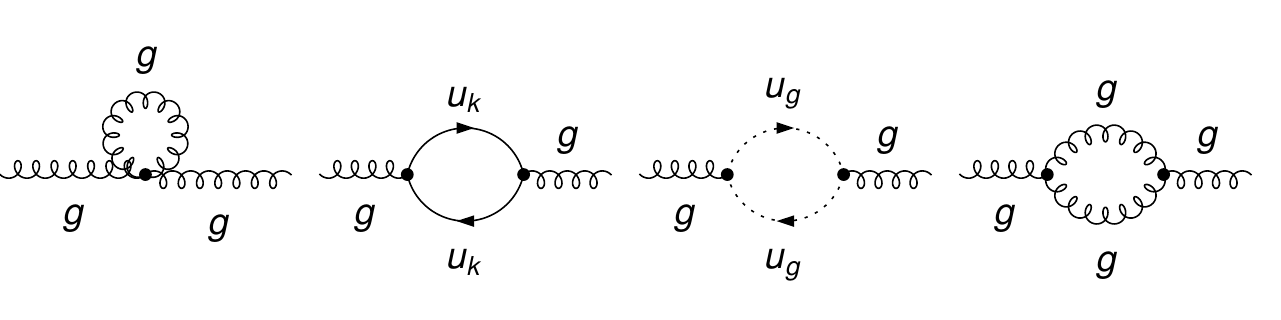}
\caption{1-loop contributions to the gluon self-energy in QCD}
\label{fig:gluonse}
\end{figure}

\begin{lstlisting}[language=Mathematica]
In[2]:= diags = InsertFields[CreateTopologies[1, 1 -> 1 , ExcludeTopologies -> {Tadpoles}],
   		{V[5]} -> {V[5]}, InsertionLevel -> {Classes},  Model -> "SMQCD",
   		ExcludeParticles -> {S[_], V[2 | 3], U[1 | 2 | 3 | 4], F[4]}];
\end{lstlisting}
With the following code 
\begin{lstlisting}[language=Mathematica,mathescape]
In[3]:= amps = FCFAConvert[CreateFeynAmp[diags, Truncated -> True, PreFactor -> -1],
    IncomingMomenta -> {p}, OutgoingMomenta -> {p}, LoopMomenta -> {q}, 
    DropSumOver -> True, ChangeDimension -> D, UndoChiralSplittings -> True,
	LorentzIndexNames -> {$\mu$, $\nu$}] /. {MQU[Index[Generation, 3]] -> m};
\end{lstlisting}
the amplitudes generated by \feynarts are converted into a valid  \feyncalc input. Here the output of the standard \feynarts function \texttt{CreateFeynAmp} is passed to \texttt{FCFAConvert}, a new \feyncalc routine that handles the conversion according to the options submitted by the user. The resulting expression $\texttt{amps}$ is a list with four entries that correspond to the four diagrams displayed in Fig. \ref{fig:gluonse}, where for convenience the \feynarts expression for the up-type quark mass \texttt{MQU[Index[Generation, 3]]} was replaced by \texttt{m}.

These four diagrams can be evaluated with
\begin{lstlisting}[language=Mathematica]
In[4]:= amps2 = TID[#, q] & /@ SUNSimplify /@ Contract /@ amps;
\end{lstlisting}
where we apply the \feyncalc functions \texttt{Contract} (contraction of Lorentz indices), \texttt{SUNSimplify} (simplification of the $SU(N)$ algebra) and \texttt{TID} (tensor reduction of 1-loop integrals) to each amplitude. We known that the first amplitude in Fig. \ref{fig:gluonse} vanishes in dimensional regularization, as it is proportional to a scaleless loop integral. \feyncalc confirms that this is indeed the case
\begin{lstlisting}[language=Mathematica]
In[5]:= amps2[[1]]
Out[5]:= 0
\end{lstlisting}
Summing up the contributions from all the diagrams (for simplicity we consider here only one quark flavor) we obtain
\begin{lstlisting}[language=Mathematica,mathescape]
In[6]:= ampsTotal = Total[amps2] // Simplify
Out[6]:= $\frac{g_s^2 \delta ^{\text{Glu1}\text{Glu2}} \left(p^{\mu } p^{\nu }-p^2 g^{\mu \nu }\right) \left(\frac{(3 D-2) p^2 C_A}{q^2.(q-p)^2}-\frac{2 \left((D-2) p^2+4 m^2\right)}{\left(q^2-m^2\right).\left((q-p)^2-m^2\right)}+\frac{4 (D-2)}{q^2-m^2}\right)}{2 (D-1) p^2}$
\end{lstlisting}
For convenience, we may also rewrite this result in terms of the Passarino-Veltman scalar functions $A_0$ and $B_0$ by applying the function \texttt{ToPaVe} that is available since \feyncalc 9

\begin{lstlisting}[language=Mathematica,mathescape]
In[7]:= res = ampsTotal // ToPaVe[#, q] & // Simplify
Out[7]:= $\frac{i \pi ^2 g_s^2 \delta ^{\text{Glu1}\text{Glu2}} \left(p^2 g^{\mu \nu }-p^{\mu } p^{\nu }\right) \left((2-3 D) p^2 C_A \text{B}_0\left(p^2,0,0\right)+2 \left((D-2) p^2+4 m^2\right) \text{B}_0\left(p^2,m^2,m^2\right)-4 (D-2) \text{A}_0\left(m^2\right)\right)}{2 (D-1) p^2}$
\end{lstlisting}
Such an output is the simplest possible form that can be achieved with \feyncalc. However, with the aid of some auxiliary tools one can in principle obtain even more explicit results with the 1-loop master integrals evaluated analytically. An add-on called \textsc{FeynHelpers} that connects \feyncalc to such tools is currently in development, with the early version of the code already available for testing \footnote{github.com/FeynCalc/feynhelpers}

\section{Summary} \label{sec:summary}

In summary, we described some of the new developments in \feyncalc 9 and provided an explicit example for using the combination of \feyncalc and \feynarts to compute the 1-loop gluon self-energy in QCD. While better support for multi-loop calculations is one of the important goals in the future development of \feyncalc, we are also looking into possibilities to combine \feyncalc with other useful software tools by developing the corresponding interfaces. 

\section*{Acknowledgments}

This work has been supported by the DFG and the NSFC through funds provided to the Sino-German CRC 110 ``Symmetries and the Emergence of Structure in QCD'', and by the DFG cluster of excellence ``Origin and structure of the universe'' (\url{www.universe-cluster.de}). This report has the preprint number TUM-EFT 78/16.

\section*{References}

\bibliographystyle{iopart-num}

\bibliography{bibsources}

\end{document}